\documentclass[12pt]{article}
\usepackage{graphicx}
\begin{document}

\title{Radiative corrections to deep--inelastic $ed-$ scattering. Case
of tensor polarized deuteron }

 \author{G. I. Gakh and O. N. Shekhovtsova}
 \date{}
 \maketitle
 \begin{center}

{\small {\it National Science Centre "Kharkov Institute of }}
{\it{Physics and Technology,\\ 61108 Akademicheskaya 1, Kharkov,
Ukraine}} \\
{\small {\it {e-mail: shekhovtsova@kipt.kharkov.ua}}}
\end{center}

\vspace{0.5cm}
\begin{abstract}
The model--independent radiative corrections to deep--inelastic
scattering of unpolarized electron beam off the tensor polarized
deuteron target have been considered. The contribution to the
radiative corrections due to the hard--photon emission from the
elastic electron--deuteron scattering (the so--called elastic
radiative tail) is also investigated. The calculation is based on
the covariant parametrization of the deuteron quadrupole
polarization tensor. The numerical estimates of the radiative
corrections to the polarization observables have been done for the
kinematical conditions of the current experiment at HERA.
\end{abstract}
\vspace{0.2cm}
\vspace{0.2cm}

\section{Introduction}
\hspace{0.7cm}

The flavor structure of nucleons is described in terms of parton
distribution functions. Most of the information on these functions has
up to now come from inclusive deep--inelastic scattering process:
experiments where only the scattered lepton is detected. The investigation
of the nucleon spin structure involves now new types of reactions. For
example, the HERMES experiment was specifically designed to perform accurate
measurements of semi--inclusive reactions, where besides scattered lepton
also some of the produced hadrons are detected
\cite{R}.

The polarized nuclei of deuterium and helium--3 are used to extract
information on the neutron spin--dependent structure function $g_1(x)$
\cite{SMC}. In analyzing the experimental data on inclusive spin
asymmetries for deuterium in order to deduce the spin--dependent structure
function $g_1^d$ one should to take into account a small effect due to possible
tensor polarization in this spin--one target. This is connected with the
presence in a deuteron target of an additional tensor polarized structure
functions \cite{R}. So far, the spin--structure studies have been focused on
the spin-1/2 nucleon. Different spin physics exists for higher--spin hadrons
such as the tensor structure in the deuteron. The measurement of these
additional spin--dependent structure functions provides important information
about non--nucleonic components in spin--one nuclei and tensor structures on
the quark--parton level \cite{K1}. A general formalism of the deep inelastic
electron--deuteron scattering was discussed in Ref. \cite{HJM}, where new
four tensor structure functions $b_i(x), \ i=1-4 $ were introduced. They can
be measured using tensor polarized target and unpolarized electron beam.
Among these new structure functions, only one, $b_1$, is leading twist in QCD
\cite{HJM} and it was found that this function is small for a weakly bound
system of nucleons (for example, deuteron). Therefore, the measurement of
$b_1$ for the case of the deuteron can give information about its possible
exotic components.

From the theoretical point of view the spin--dependent structure
function $b_1(x)$ was investigated in a number of papers. The
available fixed targets with $J\geq 1$ are only nuclei (the
deuteron is the most commonly used nucleus). If the nucleons in
the deuteron are in an $S$ state then $b_1(x)\equiv 0.$ For
nucleons in a $D$ state, $b_1(x)\not= 0$ in general \cite{HJM}. It
was found \cite{CK} that in a quark--parton model the sum rule
$\int dxb_1(x)= 0 $ is generally true if the sea of quarks and
antiquarks is unpolarized (and it was shown how this sum rule is
modified in the presence of a polarized sea). Mankiewicz \cite{M}
has studied the $b_1(x)$ for the $\rho $ meson and noticed
empirically that $\int dxb_1(x) = 0 $ in his model. It was shown
in Ref. \cite{KH} that multiple scattering terms at low $x$ can
still lead to $b_1\not= 0$ even if only the $S$--wave component is
present. Various twist two structure functions of the deuteron (in
particular, $b_1$) have been calculated in a version of the
convolution model that incorporates relativistic and binding
energy corrections \cite{KH1}. A simple parametrizations of these
structure functions are given in terms of a few deuteron wave
function parameters and the free nucleon structure functions. The
tensor structure functions were discussed in Ref. \cite{K} for the
case of lepton scattering and in hadron reactions such as the
polarized proton--deuteron Drell--Yan process.

As it is known, the HERMES experiment has been designed to measure
the nucleon spin--dependent structure functions from
deep--inelastic scattering of longitudinally polarized positrons
and electrons from polarized gaseous targets ($H, D, ^3He$). In
2000, HERMES collected a data set with a tensor polarized
deuterium target for the purpose of making a first measurement of
tensor structure function $b_1(x).$ The preliminary results on
this structure function are presented in Ref. \cite{C} for the
kinematic range $0.002<x<0.85$ and $0.1 GeV^2<Q^2<20 GeV^2.$ The
preliminary result for the tensor asymmetry is sufficiently small
to produce an effect of more than $1\% $ on the measurement of
$g_1^d.$ The dependence of $b_1 $ on $x $ variable is in
qualitative agreement with expectations based on coherent double
scattering models \cite{N, E, B} and favors a sizeable  value of
$b_1$ at low $x$ region. This suggests a significant tensor
polarization of the sea--quarks, violating the Close--Kumano sum
rule \cite{CK}.

The radiative corrections to deep--inelastic scattering of
unpolarized and longitudinally polarized electron beam on
polarized deuteron target were considered in Ref. \cite{AS} for
the particular case of the deuteron polarization (which can be
obtained from the general covariant spin--density matrix
\cite{CLS} when spin functions are the eigenvectors of the spin
projection operator). The leading--log model--independent
radiative corrections in deep--inelastic scattering of unpolarized
electron beam off the tensor polarized deuteron target have been
considered in Ref. \cite{GM}. The calculation is based on the
covariant parametrization of the deuteron quadrupole polarization
tensor and use of the Drell--Yan like representation in
electrodynamics.

Current experiments at modern accelerators reached a new level of
precision and this circumstance requires a new approach to data analysis
and inclusion of all possible systematic uncertainties. One of the
important source of such uncertainties is the electromagnetic radiative
effects caused by physical processes which take place in higher orders of
the perturbation theory with respect to the electromagnetic interaction. In
present paper we  give the covariant description of deep--inelastic scattering
of unpolarized
electron beam off the tensor polarized deuteron target (the polarization state
of the target is described by the spin--density matrix of the general form)
taking into account the radiative corrections
\begin{equation}\label{1}
e^-(k_1)+d(p)\rightarrow e^-(k_2) + X(p_x).
\end{equation}

The corresponding approach is based on the covariant parametrization of
the deuteron quadrupole polarization tensor in terms of the 4--momenta of
the particles in process (1) \cite{GM}. We performed also the numerical
calculations of the radiative corrections for the kinematical conditions
of the experiment \cite {C}. The contribution of the radiative tail from
the elastic $ed-$scattering is considered separately.

\section{Born approximation}
\hspace{0.7cm}

The standard set of variables which is usually used for the description of
deep--inelastic scattering process is
\begin{equation}\label{2}
x=\frac{-q^2}{2pq}, \ y=\frac{2pq}{V}, \ V=2pk_1, \ q^2=-Vxy, \ q=k_1-k_2,
\end{equation}
where $q$ is the 4--momentum of the intermediate heavy photon that probes
the deuteron structure.
To begin with, we define the deep--inelastic scattering cross section of the process (1) in terms
of the leptonic $L_{\mu\nu}$ and hadronic $W_{\mu\nu}$ tensors contraction
(in the Born approximation we can neglect the electron mass)
\begin{equation}\label{3}
\frac{d\sigma}{dxdQ_B^2} = \frac{\pi\alpha^2}{VQ_B^4}\frac{y}{x}
L_{\mu\nu}W_{\mu\nu}\ . \
\end{equation}
Note that only in the Born approximation (without
accounting radiative corrections) $q=k_1-k_2, \ Q_B^2=-q^2=2k_1k_2.$

The Born leptonic tensor (for the unpolarized case) is
\begin{equation}\label{4}
L^B_{\mu\nu}=q^2g_{\mu\nu}+2(k_{1\mu}k_{2\nu}+k_{1\nu}k_{2\mu}).
\end{equation}

The hadronic tensor is defined by the following way
$$W_{\mu\nu}=(2\pi )^3\sum_X \delta ^{(4)}(k_1+p-k_2-p_x)
\overline{J_{\mu }J^*_{\nu }}, $$ where $J_{\mu }$ is the
electromagnetic current for the $\gamma ^*+d \rightarrow X$
transition ($\gamma ^*$ is the virtual photon). The sum means
summation over the final states and the bar means averaging over
the polarizations of the target and summation over the
polarizations of the final particles. To write down the hadron
tensor in terms of the structure functions we define first the
deuteron spin--density matrix (further we do not consider the
effect caused by the vector polarization of the deuteron)
\begin{equation}\label{5}
\rho_{\mu\nu}=-\frac{1}{3}\bigl(g_{\mu\nu}-\frac{p_{\mu}p_{\nu}}{M^2}\bigr)
-\frac{i}{2M}\varepsilon_{\mu\nu\lambda\rho}s_{\lambda}p_{\rho}+ Q_{\mu\nu},
\ \ Q_{\mu\nu}=Q_{\nu\mu}, \ \ Q_{\mu\mu}=0\ , \ \ p_{\mu}Q_{\mu\nu}=0\ ,
\end{equation}
here $s_{\mu}$ and $Q_{\mu\nu}$ are the target--deuteron polarization
4-vector and the deuteron quadrupole--polarization tensor. The corresponding
hadron tensor has both the polarization--independent and polarization--
dependent parts and in general case can be written as
\begin{equation}\label{6}
W_{\mu\nu}=W_{\mu\nu}(0)+W_{\mu\nu}(V)+W_{\mu\nu}(T),
\end{equation}
where $W_{\mu\nu}(0)$ corresponds to the unpolarized case and
$W_{\mu\nu}(V) (W_{\mu\nu}(T))$ corresponds to the case of the
vector (tensor-)-polarization of the deuteron target. The
$W_{\mu\nu}(0)$ term has the following form
\begin{equation}\label{7}
W_{\mu\nu}(0)=-W_1\tilde{g}_{\mu\nu}+\frac{W_2}{M^2}\tilde{p}_{\mu}
\tilde{p}_{\nu} \ , \
\tilde{g}_{\mu\nu}=g_{\mu\nu}-\frac{q_{\mu}q_{\nu}}{q^2}\ , \ \
\tilde{p}_{\mu}=p_{\mu}-\frac{pq}{q^2}q_{\mu} \ ,
\end{equation}
here $M$ is the deuteron mass, and $W_{1,2}$ are the unpolarized
structure functions depending on two independent variables $x$ and
$q^2.$ The part of the hadron tensor which is dependent on the
quadrupole polarization tensor can be represented as
\begin{equation}\label{8}
W_{\mu\nu}(T)=\frac{M^2}{(pq)^2}\bigl\{Q_{\alpha\beta}q_{\alpha}q_{\beta}
\bigl(B_1\tilde{g}_{\mu\nu}+\frac{B_2}{pq}\tilde{p}_{\mu}\tilde{p}_{\nu}
\bigr)+B_3q_{\alpha}(\tilde{p}_{\mu}Q_{\tilde{\nu}\alpha}+
\tilde{p}_{\nu}Q_{\tilde{\mu}\alpha})+pqB_4\widetilde{Q}_{\mu\nu}\bigr\}.
\end{equation}
Here $B_i(i=1,2,3,4)$ are the spin--dependent structure functions
(caused by the tensor polarization of the target). They are also
functions of two variables $q^2$ and $x$. Since the hadron tensor
$W_{\mu\nu}(T)$ is symmetric under $\mu\leftrightarrow\nu$ the
electron beam does not have to be polarized for measuring these
new structure functions.

We used the following notation in formula (8)
$$Q_{\mu\tilde{\nu}}=Q_{\mu\nu}-\frac{q_{\nu}q_{\alpha}}{q^2}Q_{\mu\alpha}
\ , \ \ Q_{\mu\tilde{\nu}}q_{\nu}=0\ , $$
\begin{equation}\label{9}
\widetilde{Q}_{\mu\nu}=
Q_{\mu\nu}+\frac{q_{\mu}q_{\nu}}{q^4}Q_{\alpha\beta}q_{\alpha}q_{\beta}-
\frac{q_{\nu}q_{\alpha}}{q^2}Q_{\mu\alpha}-
\frac{q_{\mu}q_{\alpha}}{q^2}Q_{\nu\alpha}\ , \
\widetilde{Q}_{\mu\nu}q_{\nu} = 0 \ .
\end{equation}
Note that the deuteron spin--dependent structure functions $B_i$ are also
related to the structure functions $b_i$, introduced in Ref. \cite{HJM},
in the following way
\begin{equation}\label{10}
B_1=-b_1, \ B_2=\frac{b_2}{3}+b_3+b_4, \
B_3=\frac{b_2}{6}-\frac{b_4}{2}, \ B_4=\frac{b_2}{3}-b_3.
\end{equation}

When calculating radiative corrections it is convenient to
parametrize the polarization state of the deuteron target in terms
of the 4-momenta of the particles participating in the reaction
under consideration. Therefore, first, we have to find the set of
the axes and write them in covariant form in terms of the
4--momenta. If we choose, in the laboratory system of  reaction
(1), the longitudinal direction {\bf l} along the electron beam
and the transverse one {\bf t} in the plane $({\bf k_1,k_2})$ and
perpendicular to {\bf l}, then
\begin{equation}\label{11}
S^{(l)}_{\mu}=\frac{2\tau k_{1\mu}-p_{\mu}}{M}, \
S^{(t)}_{\mu}=\frac{k_{2\mu}-(1-y-2xy\tau)k_{1\mu}-xyp_{\mu}}{d}, \
S^{(n)}_{\mu}=\frac{2\varepsilon_{\mu\lambda\rho\sigma} p_{\lambda} k_{1\rho}
k_{2\sigma}}{Vd}\ ,
\end{equation}
$$d=\sqrt{Vxyb}, \ \ b =1-y-xy\tau, \ \ \tau =M^2/V. $$
We chose one of the axes along the direction {\bf l} because in the experiment
on measuring the $b_1$ structure function \cite{C} the direction of the magnetic
field, used for the polarization of the deuteron target, is along the positron
beam line. The direction of the magnetic field provides the quantization axis
for the nuclear spin in the target.

One can verify that the set of the 4-vectors $S_{\mu}^{(l,t,n)}$ satisfies
the following properties
\begin{equation}\label{12}
S_{\mu}^{(\alpha)}S_{\mu}^{(\beta)} = -\delta_{\alpha\beta}, \ \
S_{\mu}^{(\alpha)}p_{\mu} =0, \ \ \alpha,  \beta = l,t,n.
\end{equation}
One can make sure also that in the rest frame of the deuteron (the laboratory
system)
$$ S_{\mu}^{(l)}=(0,{\bf l}), \ \ S_{\mu}^{(t)}=(0,{\bf t}), \ \
S_{\mu}^{(n)}=(0,{\bf n}) \ , $$
\begin{equation}\label{13}
{\bf l = n_1}, \ \ {\bf t} = \frac{{\bf
n_2-(n_1n_2)n_1}}{\sqrt{1-{\bf(n_1n_2)}^2}}, \ \ {\bf n}=\frac{{\bf
n_1\times n_2}}{\sqrt{1-{\bf(n_1n_2)}^2}}, \ \ {\bf n_{1,2}}=
\frac{{\bf k_{1,2}}}{|{\bf k_{1,2}}|}\ .
\end{equation}

If to add one more 4--vector $S_{\mu}^{(0)}=p_{\mu}/M$ to the set (11),
we receive the complete set of the orthogonal 4--vectors with the following
properties
\begin{equation}\label{14}
S_{\mu}^{(m)}S_{\nu}^{(m)} = g_{\mu\nu}, \ \
S_{\mu}^{(m)}S_{\mu}^{(n)} = g_{mn}, \ \ m,n = 0,l,t,n.
\end{equation}
This allows to express the deuteron quadrupole polarization tensor, in
general case, as follows
\begin{equation}\label{15}
Q_{\mu\nu} = S_{\mu}^{(m)}S_{\nu}^{(n)}R_{mn} \equiv S_{\mu}^{(\alpha)}
S_{\nu}^{(\beta)}R_{\alpha\beta}, \ \ R_{\alpha\beta}=R_{\beta\alpha}, \
R_{\alpha\alpha}=0
\end{equation}
because the components $R_{00},\ R_{0\alpha}$ and $R_{\alpha 0}$
identically equal to zero due to condition $Q_{\mu\nu}p_{\nu}=0.$

In the Born approximation the components $R_{ln}$ and $R_{tn}$ do not
contribute to the cross section (since the 4-momenta $q_{\mu}$ and $k_{1\mu}$
are orthogonal to the 4-vector $S_{\mu}^{(n)}$) and the expansion (15) can be
rewritten in the following standard form
\begin{equation}\label{16}
Q_{\mu\nu}=\bigl[S_{\mu}^{(l)}S_{\nu}^{(l)}-\frac{1}{2}S_{\mu}^{(t)}
S_{\nu}^{(t)}\bigr]R_{ll} +\frac{1}{2}S_{\mu}^{(t)}S_{\nu}^{(t)}\bigl(
R_{tt}-R_{nn}\bigr)+\bigl(S_{\mu}^{(l)}S_{\nu}^{(t)}
+S_{\mu}^{(t)}S_{\nu}^{(l)}\bigr)R_{lt}\ ,
\end{equation}
here we took into account that $R_{ll}+R_{tt}+R_{nn}=0.$

In further, the deep--inelastic scattering of the unpolarized electron beam
from the tensor polarized deuteron target is considered. Thus, we have to
calculate only the convolution of the Born leptonic
tensor $L^B_{\mu\nu}$ and hadron tensor $W_{\mu\nu}(T)$ caused by the
tensor polarization of the target
\begin{equation}\label{17}
S^B(T)=L^B_{\mu\nu}W_{\mu\nu}(T)=8\frac{\tau}{y}\bigl\{-\frac{1}{y^2}
[xy^2B_1+(a-1+y)B_2+yB_3]Q_0+
\end{equation}
$$+\frac{1}{y}[(2-y)B_3-yB_4]Q_1+B_4Q_{11}\bigr\}, \ $$
where $a=xy\tau,$ $Q_0=Q_{\alpha\beta}q_{\alpha}q_{\beta},$
$Q_1=Q_{\alpha\beta}q_{\alpha}k_{1\beta},$ $Q_{11}=Q_{\alpha\beta}
k_{1\alpha}k_{1\beta}.$ Using the formulae for the vectors $S^{(\alpha )}
_{\mu}$ we can calculate the convolutions. After simple calculations we have
\begin{equation}\label{18}
\frac{d\sigma_B(T)}{dxdQ_B^2}= \frac{2\pi\alpha^2}{xQ_B^4}
\bigl[S_{ll}R_{ll}+S_{tt}(R_{tt}-R_{nn})+S_{lt}R_{lt}\bigr],
\end{equation}
with
$$S_{ll}=\bigl[2xb\tau-y(1+2x\tau )^2\bigr]G+2b(1+3x\tau )B_3+(b-a)B_4, $$
\begin{equation}\label{19}
S_{lt}=2\sqrt{\frac{xb\tau}{y}}\bigl[2(y+2a)G+(2-y-4b)B_3+yB_4\bigr],
\end{equation}
$$S_{tt}=-2xb\tau(G+B_3), \ \ G=xyB_1-\frac{b}{y}B_2\ . $$

So, in the general case, the cross section of deep--inelastic scattering of unpolarized electron
beam from the tensor polarized target is determined, in the Born approximation,
by the components of the quadrupole--polarization tensor $R_{ll}$, $R_{lt}$
and the combination $(R_{tt}-R_{nn}).$

Consider just one more, commonly used, choice of the coordinate axes:
components of the deuteron polarization tensor are defined in the coordinate
system with the axes along directions {\bf L,\ T} and {\bf N} in the rest
frame of the deuteron, where
\begin{equation}\label{20}
{\bf L = \frac{k_1-k_2}{|k_1-k_2|}}, \ \ {\bf T} = \frac{{\bf
n_1-(n_1L)L}}{\sqrt{1-{\bf(n_1L)}^2}}, \ \ {\bf N = n} \ .
\end{equation}
The corresponding covariant form of set (20) reads
\begin{equation}\label{21}
S_{\mu}^{(L)} =\frac{2\tau(k_1-k_2)_{\mu} -yp_{\mu}}{M\sqrt{yh}}\ , \
S_{\mu}^{(T)} =\frac{(1+2x\tau)k_{2\mu}-(1-y-2x\tau)k_{1\mu}
-x(2-y)p_{\mu}}{\sqrt{Vxbh}}\,
\end{equation}
$$S_{\mu}^{(N)} =S_{\mu}^{(n)}\ , \ \ h=y+4x\tau\ , $$
and the expansion of the deuteron polarization tensor is defined in full
analogy with (16)
\begin{equation}\label{22}
Q_{\mu\nu} = \bigl[S_{\mu}^{(L)}S_{\nu}^{(L)}-\frac{1}{2}
S_{\mu}^{(T)}S_{\nu}^{(T)}\bigr]R_{LL}
+\frac{1}{2}S_{\mu}^{(T)}S_{\nu}^{(T)}\bigl(R_{TT}-R_{NN}\bigr)
+\bigl(S_{\mu}^{(L)}S_{\nu}^{(T)}+ S_{\mu}^{(T)}S_{\nu}^{(L)}\bigr)
R_{LT}\ .
\end{equation}
These two sets of the orthogonal 4-vectors are connected by means of
orthogonal matrix which describes the rotation in the plane perpendicular
to direction {\bf n = N}
\begin{equation}\label{23}
S_{\mu}^{(L)} = \cos{\theta}S_{\mu}^{(l)}+\sin{\theta}S_{\mu}^{(t)}, \ \
S_{\mu}^{(T)} = -\sin{\theta}S_{\mu}^{(l)} +\cos{\theta}S_{\mu}^{(t)},
\end{equation} $$\cos{\theta}=\frac{y(1+2x\tau)}{\sqrt{yh}}\ , \ \
\sin{\theta}= -2\sqrt{\frac{xb\tau}{h}}\ . $$

The part of the differential cross section that depends on the tensor
polarization can be written as follows in this set of axes
\begin{equation}\label{24}
\frac{d\sigma_B(T)}{dxdQ_B^2}=
\frac{2\pi\alpha^2}{xQ_B^4}\bigl[S_{LL}R_{LL}+S_{TT}(R_{TT}-R_{NN})+
S_{LT}R_{LT}\bigr]\ ,
\end{equation}
$$S_{LL}=-hG+2bB_3+\frac{B_4}{h}[(1-y)(y-2x\tau)-2a(y+x\tau)]\ ,$$
\begin{equation}\label{25}
S_{TT}=\frac{2xb\tau}{h}B_4 , \ \ S_{LT}=2\sqrt{\frac{xb\tau}{y}}(2-y)
\bigl(B_3+\frac{y}{h}B_4\bigr) \ .
\end{equation}

\section{Radiative corrections}
\hspace{0.7cm}

In this work we consider only QED radiative corrections to the
deep--inelastic scattering process (1). We confine ourselves to
the calculation of the so--called model--independent radiative
corrections when the photons are radiated from the lepton line and
the vacuum polarization is also taken into account. The reason is
that it gives the main contribution to radiative corrections due
to the smallness of the electron mass, and can be calculated
without any additional assumptions. Nevertheless, these radiative
corrections depend on the shape of the deuteron structure
functions (both spin--independent
 and spin--dependent) by their dependence on the $x$ and $Q^2$ variables.

There exist two contributions for radiative corrections when we
take into account the corrections of the order of $\alpha$. The
first one is caused by virtual and soft photon emission that
cannot affect the kinematics of process (1). The second one arises
due to the radiation of a hard photon
\begin{equation}\label{26}
e^-(k_1)+d(p)\rightarrow e^-(k_2)+\gamma(k)+X(p_x)\ .
\end{equation}

The leptonic tensor, corresponding to the hard--photon radiation, is well
known \cite{KMF,SK}. For the case of unpolarized electron beam it can be
written as
\begin{equation}\label{27}
L^{\gamma}_{\mu\nu}=A_0\widetilde{g}_{\mu\nu}+A_1\widetilde{k}_{1\mu}
\widetilde{k}_{1\nu}+A_2\widetilde{k}_{2\mu}\widetilde{k}_{2\nu}\ ,
\end{equation}
where
$$A_0=-\frac{(q^2+\chi_1)^2+(q^2-\chi_2)^2}{\chi_1\chi_2}-
2m^2q^2\bigg(\frac{1}{\chi_1^2}+\frac{1}{\chi_2^2}\bigg), \
A_1=-4\bigg(\frac{q^2}{\chi_1\chi_2}+\frac{2m^2}{\chi_2^2}\bigg), \ $$
$$A_2=-4\bigg(\frac{q^2}{\chi_1\chi_2}+\frac{2m^2}{\chi_1^2}\bigg), \
\tilde{k}_{i\mu}=k_{i\mu}-\frac{qk_i}{q^2}q_{\mu} \ , (i=1,2), \
$$ with $\chi_{1,2}=2kk_{1,2}, $ $m$ is the electron mass,
$q^2=\chi_2-\chi_1- Q_B^2,$ and in this chapter $q=k_1-k_2-k$. The
hadronic tensor in this case has the same form as the hadronic
tensor in the Born approximation, but momentum transfer $q$
differs from the Born one and the structure functions $B_i$ depend
on the new momentum $q$. Here and further we neglect the terms
which are zero at $m\rightarrow 0.$

We consider the hard--photon (with the energy $\omega >\Delta
\varepsilon$, where $\Delta <<1$) emission process using the
approach of paper \cite{KMF1} where it was applied to the process
of deep--inelastic scattering on unpolarized target. Let us
introduce the variables suitable for this process
$$z=\frac{M_x^2-M^2}{V}=\frac{q^2+2pq}{V}, \ r=-\frac{q^2}{Q_B^2}, \
x'=\frac{-q^2}{2pq}=\frac{xyr}{xyr+z}, \ \chi_{1,2}=2kk_{1,2}, $$
where $M_x$ is the invariant mass of the hadron system produced in the
scattering of the photon (with the virtuality $q^2$) by the target.

Note that the physical meaning of $z$ variable is following: the quantity $z$
shows the degree of deviation from the elastic process ($ed\rightarrow ed$).
So, the value $z=0$ corresponds to the elastic $ed$--scattering threshold and
the value $z=\varepsilon_d/\varepsilon_1 $ ($\varepsilon_d$ is the deuteron
bound energy, $\varepsilon_1$ is the electron beam energy in the laboratory
system) corresponds to the $ed\rightarrow enp$ reaction threshold
(quasi--elastic $ed-$ scattering).

The convolution of the leptonic and hadronic tensors may be represented as
\begin{equation}\label{28}
S^{\gamma}(T)=AA_0+BA_1+CA_2, \
\end{equation}
$$A=NQ_0\bigl[3B_1+\frac{2\tau }{c}B_2+
\frac{c}{2xyr}(B_2+2B_3+B_4)\bigr], $$
$$B=N\bigl\{Q_0\bigl[\frac{V}{2c}B_2-
V\frac{Q_B^2+\chi_1 }{2rQ_B^2}(B_2+B_3)+\frac{(Q_B^2+\chi_1)^2}{4rQ_B^2}(B_1+ $$
$$+\frac{Vc}{2rQ_B^2}(B_2+2B_3+B_4))\bigr]+VQ_1\bigl[B_3-
\frac{Q_B^2+\chi_1}{2rQ_B^2}c(B_3+B_4)\bigr]+
\frac{V}{2}cQ_{11}B_4\bigr\},  $$
$$C=N\bigl\{Q_0\bigl[\frac{V}{2}\frac{(1-y)^2}
{c}B_2+V\frac{Q_B^2-\chi_2 }{2rQ_B^2}(1-y)(B_2+B_3)+
\frac{(Q_B^2-\chi_2)^2}{4rQ_B^2}(B_1+ $$
$$+\frac{Vc}{2rQ_B^2}(B_2+2B_3+B_4))\bigr]+VQ_2\bigl[B_3(1-y)+
\frac{Q_B^2-\chi_2}{2rQ_B^2}c(B_3+B_4)\bigr]+
\frac{V}{2}cQ_{22}B_4\bigr\},  $$
with $N=4\tau /Vc^2, \ c=z+xyr.$

The quantities $Q_0, \ Q_1, \ Q_2, \ Q_{11}$ and $Q_{22}$ are the convolutions
of the deuteron quadrupole--polarization tensor and 4-momenta and they can be
expressed in terms of the scalar products of the 4-momenta of the particles
participating in the reaction and the set of 4-vectors $S_{\mu}^{(l,t,n)}$.
So, these convolutions are
$$Q_0=Q_{\alpha\beta}q_{\alpha}q_{\beta}=[(l\cdot q)^2-\frac{1}{2}(t\cdot q)^2-
\frac{1}{2}(n\cdot q)^2]R_{ll}+2l\cdot qt\cdot qR_{lt}+2n\cdot ql\cdot qR_{ln}+ $$
$$+2n\cdot qt\cdot qR_{tn}+\frac{1}{2}[(t\cdot q)^2-(n\cdot q)^2](R_{tt}-R_{nn}), $$
$$Q_1=Q_{\alpha\beta}q_{\alpha}k_{1\beta}=
(l\cdot k_1l\cdot q-\frac{1}{2}t\cdot k_1t\cdot q)R_{ll}+(l\cdot k_1t\cdot q
+t\cdot k_1l\cdot q)R_{lt}+l\cdot k_1n\cdot qR_{ln}+ $$
\begin{equation}\label{29}
+t\cdot k_1n\cdot qR_{tn}+\frac{1}{2}t\cdot k_1t\cdot q(R_{tt}-R_{nn}),
\end{equation}
$$Q_{11}=Q_{\alpha\beta}k_{1\alpha}k_{1\beta}=
[(l\cdot k_1)^2-\frac{1}{2}(t\cdot k_1)^2]R_{ll}+2l\cdot k_1t\cdot k_1R_{lt}+
\frac{1}{2}(t\cdot k_1)^2(R_{tt}-R_{nn}), $$
$$Q_2=Q_1(k_1\rightarrow k_2), \ Q_{22}=Q_{11}(k_1\rightarrow k_2), \
i\cdot a=S_{\mu }^{(i)}a_{\mu }, \ i=l,t,n. $$
Here we used the following conditions: $R_{ll}+R_{tt}+R_{nn}=0, \ n\cdot k_1=
n\cdot k_2=0.$ For the set of the 4-vectors $S_{\mu }^{(l,t,n)}$ we have also
$t\cdot k_1=0.$

It is convenient to separate the poles in the term $(\chi_1\chi_2)^{-1}$
using the relation
$$\frac{1}{\chi_1\chi_2 }=\frac{1}{Q_B^2}\frac{1}{1-r}(\frac{1}{\chi_1 }-
\frac{1}{\chi_2 }). $$

Then the radiative correction (caused by the hard--photon emission) to the
differential cross section of deep--inelastic scattering of unpolarized
electron beam by the tensor polarized target has the following form
\begin{equation}\label{30}
\frac{d\sigma^{\gamma}}{dxdQ_B^2}=
\frac{\alpha y}{Vx}\int \frac{d^3{\vec k}}{2\pi\omega}\Sigma(z,r) ,
\end{equation}
where $\omega $ is the energy of the hard photon and
\begin{equation}\label{31}
\Sigma (z,r)=\frac{\alpha ^2(q^2)}{Q_B^4}\bigl\{ R_0(z,r)+
(\frac{1}{\chi_1 }-\frac{1}{\chi_2 })R_1(z,r)+
\frac{m^2}{\chi_1^2}R_{1m}(z,r)+
\frac{m^2}{\chi_2^2}R_{2m}(z,r)\bigr\},
\end{equation}
$$R_0=-\frac{2}{r^2}A, \ R_1=\frac{1}{r-1}[(1+\frac{1}{r^2})Q_B^2A-
\frac{4}{r}(B+C)], \ R_{1m}=2(\frac{Q_B^2}{r}A-\frac{4}{r^2}C), \
R_{2m}=2(\frac{Q_B^2}{r}A-\frac{4}{r^2}B). \ $$
It is convenient to write down the integral in Eq. (30) as
\begin{equation}\label{32}
I=\int \frac{d^3{\vec k}}{2\pi\omega}\Sigma(z,r)=I_{1m}+I_{2m}+I_R ,
\end{equation}
where we separate the contributions proportional to $m^2$
\begin{equation}\label{33}
I_{1m}=\int \frac{d^3{\vec k}}{2\pi\omega}\frac{\alpha^2(q^2)}{Q_B^4}
\frac{m^2}{\chi_1^2}R_{1m}(z,r) , \
I_{2m}=\int \frac{d^3{\vec k}}{2\pi\omega}\frac{\alpha^2(q^2)}{Q_B^4}
\frac{m^2}{\chi_2^2}R_{2m}(z,r) . \
\end{equation}

Let us consider first the integrals $I_{im}, \ i=1,2.$ In this
case the numerator of the integrands in $I_{1m}(I_{2m})$ is
calculated in the approximation $\chi_1=0 (\chi_2=0)$ \cite{KMF1}.
The hard--photon phase--space is written as
\begin{equation}\label{34}
\frac{d^3{\vec k}}{2\pi\omega}=\frac{dz}{z_+-z}
\frac{\omega^2d\Omega_k}{2\pi}, \ \  z_+=y(1-x).
\end{equation}

Using the invariance of $\omega^2d\Omega_k, $we can do the integration
over the angular variables $d\Omega_k$ in the most suitable coordinate
system, namely: in the coordinate frame ${\vec k_1}-{\vec k_2}+{\vec p}=0$
(c.m.s. of the scattered electron and produced hadronic system). We obtain
$$\int \frac{\omega^2d\Omega_k}{2\pi}\frac{m^2}{\chi_{1,2}^2}=
\frac{1}{2}. \  $$ \underbar{Integral $I_{1m}$.} We calculate the
integrand in the approximation $\chi_1=0$ (besides the
denominator). This approximation corresponds to the emission of
the collinear photon along the initial--electron momentum. In this
case the variables acquire the following values
$$r_1=\frac{1-y+z}{1-xy}, \ q_1^2=-r_1Q_B^2, \ x'_1=\frac{xyr_1}{z+xyr_1}. $$
After integrating over the hard--photon angular variables, the integral
$I_{1m}$ can be represented as follows
\begin{equation}\label{35}
I_{1m}=\frac{1}{Q_B^4}\int_0^{z_m}\frac{dz}{z_+-z}\alpha_1^2N_1\Sigma_1(z),
\end{equation}
$$\Sigma_1(z)=\Sigma_{1ll}R_{ll}+\Sigma_{1lt}R_{lt}+
\Sigma_{1tt}(R_{tt}-R_{nn}), \ $$
$$\Sigma_{1tt}=b\frac{Q_B^2}{r_1}(G_t+B_{3t}), \
\Sigma_{1lt}=-\frac{V}{r_1}\sqrt {\frac{xyb}{\tau }}\bigl [(y-1+r_1)B_{4t}+
(a-3b+r_1)B_{3t}+2(a-b+r_1)G_t\bigr ], \ $$
$$\Sigma_{1ll}=\frac{V}{2\tau r_1}\bigl\{(a-b)(y-1+r_1)B_{4t}+
2b(b-2a-r_1)B_{3t}-[2ab-(a-b+r_1)^2]G_t\bigr\}, \ $$
$$G_t=xyB_{1t}-\frac{b}{y-1+r_1}B_{2t}, \ \alpha_1=\alpha (q_1^2), \
N_1=\frac{4\tau }{(z+xyr_1)^2}, \ $$
$$z_m=z_+-\rho, \ \rho =2\Delta\varepsilon\sqrt{(\tau +z_+)/V}, \
B_{it}=B_i(q_1^2, x'_1), \ i=1-4. $$
It is convenient to extract explicitly the contribution containing the
infrared divergence. To do this we should add and subtract in the numerator of
the integrand its value at $z=z_+.$ At this value we have: $r_1=1, \ $
$\alpha_1=\alpha, \ $ $N_1=4\tau /y^2, \ $ $x'_1=x.$ So, the integral
$I_{1m}$ can be written as
\begin{equation}\label{36}
I_{1m}=\frac{1}{Q_B^4}\int_0^{z_+}\frac{dz}{z_+-z}
\Bigl [\alpha_1^2N_1\Sigma_1(z)-\alpha^2\frac{4\tau }{y^2}\Sigma_1(z_+)
\Bigr ]+\frac{Vx}{\pi y}ln(\frac{\rho }{z_+})\frac{d\sigma_B}{dxdQ_B^2}.
\end{equation}

\underbar{Integral $I_{2m}$.} Calculation of the integrand is
performed in the approximation $\chi_2=0$ that corresponds to the
emission of the collinear photon along the final--electron
momentum. In this case the variables acquire the following values
$$r_2=\frac{1-z}{1-z_+}, \ q_2^2=-r_2Q_B^2, \ x'_2=\frac{xyr_2}{1-r_2(1-y)}. $$
After integrating over the hard--photon angular variables, the integral
$I_{2m}$ is represented as follows
\begin{equation}\label{37}
I_{2m}=\frac{1}{Q_B^4}\int_0^{z_m}\frac{dz}{z_+-z}\alpha_2^2N_2\Sigma_2(z),
\end{equation}
$$\Sigma_2(z)=\Sigma_{2ll}R_{ll}+\Sigma_{2lt}R_{lt}+
\Sigma_{2tt}(R_{tt}-R_{nn}), \ $$
$$\Sigma_{2tt}=bQ_B^2(r_2G_s+B_{3s}), \
\Sigma_{2lt}=-V\sqrt {\frac{xyb}{\tau }}\bigl\{(y-1+\frac{1}{r_2})B_{4s}+
(a-3b+\frac{1}{r_2})B_{3s}+2[1+(a-b)r_2]G_s\bigr\}, \ $$
$$\Sigma_{2ll}=\frac{V}{2\tau r_2}\bigl\{(a-b)[1-r_2(1-y)]B_{4s}-
2b[1+(2a-b)r_2]B_{3s}-[2abr_2^2-(1+ar_2-br_2)^2]G_s\bigr\}, \ $$
$$G_s=xyB_{1s}-\frac{b}{1-r_2(1-y)}B_{2s}, \ \alpha_2=\alpha (q_2^2), \
N_2=\frac{4\tau }{(z+xyr_2)^2}, \ B_{is}=B_i(q_2^2, x'_2), \ i=1-4. $$
The contribution containing the infrared divergence is extracted explicitly
in a similar manner as it was done for the $I_{1m}$ integral. At the value
$z=z_+$ we have: $r_2=1, \ $ $\alpha_2=\alpha, \ $ $N_2=4\tau /y^2, \ $
$x'_2=x.$ So, the integral $I_{2m}$ is rewritten as
\begin{equation}\label{38}
I_{2m}=\frac{1}{Q_B^4}\int_0^{z_+}\frac{dz}{z_+-z}
\Bigl [\alpha_2^2N_2\Sigma_2(z)-\alpha^2\frac{4\tau }{y^2}\Sigma_2(z_+)
\Bigr ]+\frac{Vx}{\pi y}ln(\frac{\rho }{z_+})\frac{d\sigma_B}{dxdQ_B^2}.
\end{equation}

The radiative corrections due to the virtual photon exchange and real soft--
photon emission (with energy $\omega <\Delta\varepsilon $) can be related to
the Born cross section as
(note that the vacuum polarization effects are included in the Born cross--
section through the dependence of the coupling constant $\alpha $ on the
virtual--photon momentum)
\begin{equation}\label{39}
\frac{d\sigma^{(S+V)}}{dxdQ_B^2}=\delta^{SV}
\frac{d\sigma_B}{dxdQ_B^2},
\end{equation}
where the factor $\delta^{SV}$ is \cite{KMF1}
\begin{equation}\label{40}
\delta^{SV}=\frac{\alpha }{\pi }\bigl [(L-1)ln\frac{(\Delta\varepsilon )^2}
{\varepsilon_1\varepsilon_2}+\frac{3}{2}L-\frac{1}{2}ln^2
(\frac{\varepsilon_1}{\varepsilon_2})-\frac{\pi^2}{6}-2-
f(cos^2\frac{\theta }{2})\bigr ], \ L=ln\frac{Q_B^2}{m^2},
\end{equation}
here $\varepsilon_1(\varepsilon_2)$ is the initial (final) electron energy,
and $\theta $ is the electron scattering angle in the coordinate frame
${\vec k_1}-{\vec k_2}+{\vec p}=0$. The function $f$ is defined as
$$f(x)=\int_0^x \frac{dt}{t}ln(1-t). $$
The quantities $\varepsilon_1(\varepsilon_2)$ and $\theta $ can be expressed
in terms of the invariant variables
\begin{equation}\label{41}
\varepsilon_1=\frac{V(1-xy)}{2\sqrt{V(\tau +z_+)}}, \
\varepsilon_2=\frac{V(1-z_+)}{2\sqrt{V(\tau +z_+)}}, \
cos^2\frac{\theta }{2}=\frac{1-y-xy\tau }{(1-xy)(1-z_+)}.
\end{equation}
Then the radiative correction $\delta^{SV}$ is finally rewritten as
\begin{equation}\label{42}
\delta^{SV}=\frac{\alpha }{2\pi }\Bigl\{ -1-\frac{\pi^2}{3}-
2f[\frac{1-y-xy\tau }{(1-xy)(1-z_+)}]-ln^2(\frac{1-xy}{1-z_+})+
(L-1)(3+2ln[\frac{\rho^2}{(1-xy)(1-z_+)}])\Bigr\}.
\end{equation}

\underbar{Integral $I_R$.} When calculating this integral we use
the results of Ref. \cite{KMF1}. Besides the integrals calculated
in that paper we need the following integrals
\begin{equation}\label{43}
\int \frac{d^3 {\vec k}}{2\pi\omega }F(z,r)\chi _1, \
\int \frac{d^3 {\vec k}}{2\pi\omega }F(z,r)\chi _1^2. \
\end{equation}
To calculate these integrals we write the hard--photon phase space
in the form

\begin{equation}\label{44}
 \frac{d^3 {\vec k}}{2\pi\omega }=\frac{Q_B^2}{2\sqrt{y^2+4a}}
\frac{d\varphi }{2\pi }dzdr.
\end{equation}

Since, in our case, the function $F$ does not depend on the $\varphi $
variable, we can integrate over this variable. We do it in the above mentioned
coordinate frame.
The results are
\begin{equation}\label{45}
i_1=\int \frac{d\varphi }{2\pi }\chi _1=
\frac{Q_B^2}{y^2+4a}\Bigl [(2-y)(y-c)-(1-r)(y+2a)\Bigr ],
\end{equation}
$$i_2=\int \frac{d\varphi }{2\pi }\chi _1^2=
\frac{1}{2}\Bigl [3i_1^2-\frac{Q_B^4(1-xy)^2}{y^2+4a}(r-r_1)^2\Bigr ]. $$
After simple calculations the integral $I_R$ is (here we omit the contributions
proportional to the $R_{ln}$ and $R_{tn}$ components)
$$I_R=\frac{1}{2Q_B^4}\sum_{i=1}^4\sum_{m,n}R_{mn}\Bigl\{
\frac{L_1}{1-xy}\int_0^{z_m}\frac{dz}{1-r_1}G_i^{mn}(z,r_1)+
\frac{L_2}{1-z_+}\int_0^{z_m}\frac{dz}{1-r_2}\tilde G_i^{mn}(z,r_2)+   $$
\begin{equation}\label{46}
+\frac{1}{1-xy}\int_0^{z_m}dz\int_{r_-}^{r_+}\frac{dr}{|r-r_1|}
\Bigl [\frac{G_i^{mn}(z,r)}{1-r}-\frac{G_i^{mn}(z,r_1)}{1-r_1}\Bigr ]+
\end{equation}
$$+\frac{1}{1-z_+}\int_0^{z_m}dz\int_{r_-}^{r_+}\frac{dr}{|r-r_2|}
\Bigl [\frac{\tilde G_i^{mn}(z,r)}{1-r}-
\frac{\tilde G_i^{mn}(z,r_2)}{1-r_2}\Bigr ]+ $$
$$+\frac{Q_B^2}{\sqrt{y^2+4a}}\int_0^{z_m}dz\int_{r_-}^{r_+}dr
\frac{\alpha^2}{r^2}B_i\Bigl [C_{0i}^{mn}(z,r)
+i_1C_{1i}^{mn}(z,r)+i_2C_{2i}^{mn}(z,r)\Bigr ]\Bigr\}, $$
where
\begin{equation}\label{47}
L_1=ln\Bigl [\frac{Q_B^2(1-xy)^2}{m^2xy(\tau +z_+)}\Bigr ], \
L_2=ln\Bigl [\frac{Q_B^2(1-z_+)^2}{m^2xy(\tau +z_+)}\Bigr ], \
\end{equation}
$$r_{\pm }(z)=\frac{1}{2xy(\tau +z_+)}\Bigl [
2xy(\tau +z)+(z_+-z)(y\pm \sqrt{y^2+4a})\Bigr ], $$
$$G_i^{mn}(z,r)=\frac{\alpha^2}{r^2}(1-r)B_iA_i^{mn}(z,r), \
\tilde G_i^{mn}(z,r)=\frac{\alpha^2}{r^2}(1-r)B_iB_i^{mn}(z,r), \
$$ with $m,n=l,t,n.$ Note that the structure functions $B_i$ are the
functions of two independent variables $q^2=-rQ_B^2$ and
$x'=xyr/(z+xyr).$ The expressions for the coefficients $A_i^{mn},
\ $ $B_i^{mn}, \ $ $C_{ki}^{mn}, \ k=0, 1, 2$ are given in the
Appendix A. The contributions proportional to the $R_{ln}$ and
$R_{tn}$ components are considered in a more detail in the
Appendix B.

Let us discuss briefly the singularities in the $I_R$ integral.
The value $r=1$ corresponds to the real soft--photon emission (we have
infrared divergence at this point), and the value $r=r_1(r_2)$ corresponds
to the emission of the collinear photon along the initial--(final--)
electron momentum (the so--called collinear divergence). The singularity at
the point $z=z_+$ is the infrared one. The divergence at $r=1$ point is
unphysical one. It arises during the integration procedure due to the
separation of the poles in the expression $(\chi_1\chi_2)^{-1}.$ It is
necessary to extract explicitly the collinear and infrared divergences in
the above formula.

The integrand in the above expression can be written in the form
which does not contain explicitly the infrared divergences if we
add term (39) to it. To do this we use the following
transformations
\begin{equation}\label{48}
ln\Bigl [\frac{\varphi_i(x,y)}{xy(\tau +z_+)}\Bigr ]\frac{G(z,r_i)}{1-r_i}+
\int_{r_-}^{r_+}\frac{dr}{|r-r_i|}\Bigl [\frac{G(z,r)}{1-r}-
\frac{G(z,r_i)}{1-r_i}\Bigr ]=
\end{equation}
$$=\it P\int_{r_-}^{r_+}\frac{dr}{(1-r)|r-r_i|}\Bigl [G(z,r)-
G(z,r_i)\Bigr ], \ i=1,2, $$ where $\varphi_1(x,y)=(1-xy)^2, \ $
$\varphi_2(x,y)=(1-z_+)^2, \ $ and the symbol $\it P$ denotes the
principal value of integral. The total radiative correction (which
is the sum of the contribution due to the hard--photon emission
and the contribution due to the real soft--photon emission and
virtual--photon contribution) to the part of the differential
cross section caused by the tensor polarization of the target is
written as
\begin{equation}\label{49}
\frac{d\sigma }{dxdQ_B^2}=\frac{d\sigma_B}{dxdQ_B^2}+\delta^{tot},
\end{equation}
where
$$\delta^{tot}=\frac{\alpha }{2\pi }\Bigl\{3L+2(L-1)ln\Bigl [\frac{z_+^2}
{(1-xy)(1-z_+)}\Bigr ]-ln^2\Bigl (\frac{1-xy}{1-z_+}\Bigr )-4-
\frac{\pi^2}{3}- $$
$$-2f\Bigl [\frac{b}{(1-xy)(1-z_+)}\Bigr ]\Bigr\}
\frac{d\sigma_B}{dxdQ_B^2}+\frac{\alpha y}{xQ_B^4}
\int_0^{z_+}\frac{dz}{z_+-z}\Bigl [\alpha_1^2N_1\Sigma_1(z)+
\alpha_2^2N_2\Sigma_2(z)-\alpha^2\frac{8\tau }{Vy^2}\Sigma_1(z_+)\Bigr ]+  $$
\begin{equation}\label{50}
+\frac{\alpha y}{2xVQ_B^4}\sum_{i=1}^4\sum_{mn}\Bigl\{L
\int_0^{z_+}\frac{dz}{z_+-z}\Bigl [G_i^{mn}(z,r_1)-G_i^{mn}(z_+,1)-
\tilde G_i^{mn}(z,r_2)+\tilde G_i^{mn}(z_+,1)\Bigr ]+
\end{equation}
$$+\frac{Q_B^2}{\sqrt{y^2+4a}}\int_0^{z_+}dz\int_{r_-}^{r_+}dr
\frac{\alpha^2}{r^2}B_i\Bigl [C_{0i}^{mn}(z,r)+
i_1C_{1i}^{mn}(z,r)+i_2C_{2i}^{mn}(z,r)\Bigr ]+R_i^{mn}\Bigr\}. $$
The term $R_i^{mn}$ has different form depending on the integration region
of the variable $r.$ For the regions $r_-\le r\le r_1$ and $r_2\le r \le r_+$
the function $R_i^{mn}$ has the following form (in these regions $r\not= 1$ and
therefore the divergence at the point $r=1$ is absent)
\begin{equation}\label{51}
R_i^{mn}=\frac{1}{1-xy}\int_0^{z_+}dz\int_{r_-}^{r_+}
\frac{dr}{(1-r)|r-r_1|}\Bigl [G_i^{mn}(z,r)-G_i^{mn}(z,r_1)\Bigr ]+
\end{equation}
$$+\frac{1}{1-z_+}\int_0^{z_+}dz\int_{r_-}^{r_+}
\frac{dr}{(1-r)|r-r_2|}\Bigl [\tilde G_i^{mn}(z,r)-
\tilde G_i^{mn}(z,r_2)\Bigr ].  $$
For the region $r_1< r <r_2 $ we have
$$R_i^{mn}=\int_0^{z_+}dzln\Bigl (\frac{1-r_-}{r_+-1}\Bigr )\Bigl\{
g_{i1}^{mn}(z,1)-f_{i1}^{mn}(z,1)+\frac{1}{z_+-z}\Bigl [
g_{i0}^{mn}(z,1)-g_{i0}^{mn}(z,r_1)+  $$
\begin{equation}\label{52}
+f_{i0}^{mn}(z,1)-f_{i0}^{mn}(z,r_2)\Bigr ]\Bigr\}+
\int_0^{z_+}dz\int_{r_-}^{r_+}\frac{dr}{1-r}\Bigl\{
g_{i1}^{mn}(z,r)-g_{i1}^{mn}(z,1)-
\end{equation}
$$-f_{i1}^{mn}(z,r)+f_{i1}^{mn}(z,1)+\frac{1}{1-xy}\Bigl [
F^{mn}(z,r)-F^{mn}(z,1)\Bigr ]-\frac{1}{1-z_+}
\Bigl [\tilde F^{mn}(z,r)-\tilde F^{mn}(z,1)\Bigr ]\Bigr\},  $$
where we introduce the following notation
$$ G_i^{mn}(z,r)=g_{i0}^{mn}(z,r)+\Delta_1g_{i1}^{mn}(z,r), \
\tilde G_i^{mn}(z,r)=f_{i0}^{mn}(z,r)+\Delta_2 f_{i1}^{mn}(z,r), \  $$
\begin{equation}\label{53}
F^{mn}(z,r)=\frac{1}{r-r_1}\Bigl [g_{i0}^{mn}(z,r)-
g_{i0}^{mn}(z,r_1)\Bigr ], \
\tilde F^{mn}(z,r)=\frac{1}{r-r_2}\Bigl [f_{i0}^{mn}(z,r)-
f_{i0}^{mn}(z,r_2)\Bigr ], \
\end{equation}
$$\Delta_1=(1-xy)r-a-b-z, \ \Delta_2=(1-y+xy)r+z-1. $$
In obtaining the above formula we use the relation
\begin{equation}\label{54}
\it P\int_{r_-}^{r_+}\frac{dr}{1-r}\Psi (r)=
\int_{r_-}^{r_+}\frac{dr}{1-r}\Bigl [\Psi (r)-\Psi (1)\Bigr ]+
\Psi (1)ln\Bigl (\frac{1-r_-}{r_+-1}\Bigr ).
\end{equation}

At last, let us consider the part of the integral $I$ which is caused by the
$R_{ln}$ and $R_{tn}$ components of the deuteron quadrupole--polarization
tensor. As stated above, these components do not contribute to the cross
section treated in the Born approximation. If these terms are integrated
over the whole region of the $\varphi $ variable, then these integrals
are equal to zero as well (this result is due to the fact that only one plane
is remained after such integration). We discuss this problem in more detail
in Appendix B.

Note that the integration limits in formula (50) over the variable
$z$ are given somewhat schematically. This integral contains two
contributions (we neglect here the contribution of the radiative
tail from the quasi--elastic scattering). One of them is the
so--called inelastic contribution and the integration region for
it over the variables $r$ and $z$ is presented in Fig. 1 by the
dashed triangle. The integration over $z$ variable for this
contribution must be carried out from $z_{min}=(M_{th}^2-M)/V$ to
$z_+,$ where $M_{th}$ is the inelastic threshold ($M_{th}=M+m_{\pi
}$). The second contribution, related to the radiative tail of the
elastic peak, is given by the interval $z=0, \ r_-(0)\le r\le
r_+(0).$

\begin{figure}[ht]
\includegraphics[width = 0.6\textwidth, height = 0.6\textwidth]{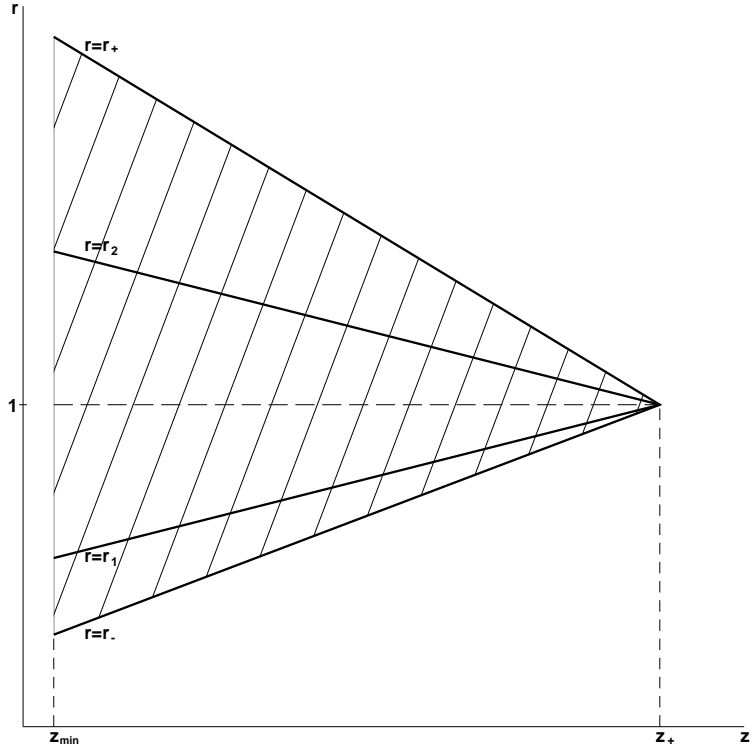}
\caption{{The integration domain with respect to the r and z
variables.}}
\end{figure}

The contribution of the elastic radiative tail (i.e. the inclusion
of radiative corrections to the elastic $ed$--scattering) to the
total radiative correction $\delta^{tot}$ can be obtained from
formula (50) by simple substitution in the hadronic tensor
\begin{equation}\label{55}
B_i(q^2,x')\rightarrow -\frac{1}{q^2}\delta (1-x')B_i^{(el)}, \ i=1-4,
\end{equation}
where $B_i^{(el)}$ are expressed in terms of the deuteron electromagnetic form
factors as
\begin{equation}\label{56}
B_1^{(el)}=\eta q^2G_M^2, \ B_2^{(el)}=-2\eta ^2q^2[G_M^2+\frac{4G_Q}{1+\eta }
(G_C+\frac{\eta }{3}G_Q+\eta G_M)],
\end{equation}
$$B_3^{(el)}=2\eta ^2q^2G_M(G_M+2G_Q), \ B_4^{(el)}=-2\eta q^2(1+\eta )G_M^2,
\ \eta =-q^2/4M^2. $$ Here $G_C, \ G_M \ $ and $G_Q$ are the
deuteron charge monopole, magnetic dipole and quadrupole form
factors, respectively. These form factors have the following
normalizations:
$$G_C(0)=1, \ G_M(0)=(M/m_n)\mu _d, \ G_Q(0)=M^2Q_d, $$
where $m_n$ is the nucleon mass, $\mu _d(Q_d)$ is the deuteron
magnetic (quadrupole) moment and their values are: $\mu _d=0.857,
Q_d=0.2859 fm^2.$ After substitution of $B_i^{(el)}$ into formula
(50) we have to do a trivial integration over $z$ variable using a
$\delta $-function $\delta (1-x')=xyr\delta (z).$

\section{Numerical estimations}
\hspace{0.7cm}

We calculate the radiative corrections for the kinematical
conditions of the HERMES experiment \cite{C}. The energy of the
positron beam is 27.6 GeV. The HERMES has provided the first
direct measurement of the structure function $b_1$ in the
kinematic range $0.002<x<0.85$ and $0.1 GeV^2<Q^2<20 GeV^2.$ A
cylindrical target cell confines the polarized gas along the
positron beam line, where a longitudinal magnetic field provides
the quantization axis for the nuclear spin. The corresponding
tensor atomic polarization is T=0.83 (for the definition of this
quantity see the Appendix C).

The analysis of the experimental data was performed in the approximation
$b_3=b_4=0.$ In further numerical estimation we also neglect these functions.

The deuteron spin--dependent structure function $b_1$ is extracted from the
measured tensor asymmetry $A_{zz}$ via the relation \cite{C}
\begin{equation}\label{57}
b_1=-\frac{3}{2}A_{zz}\frac{(1+\gamma ^2)F_2^d}{2x(1+R)},
\end{equation}
where the deuteron spin--independent structure function $F_1^d$
has been expressed in terms of the ratio $R=\sigma _L/\sigma _T=
F_2^d(1+4M^2x^2/Q^2)/2xF_1^d-1$ \cite{W}, $\gamma $ is a kinematic
factor ($\gamma ^2=4M^2x^2/Q^2$). Here $\sigma _T(\sigma _L)$ is
the cross section for the absorption of the transversely
(longitudinally) polarized virtual photons by the unpolarized
target. The Born cross section of the deep--inelastic scattering
of unpolarized electron beam by unpolarized target has the form
\begin{equation}\label{58}
\frac{d\sigma _B^{un}}{dxdQ_B^2}=\frac{4\pi\alpha ^2}{xQ_B^4}
[(1-y-xy\tau )F_2^d(x,Q^2)+xy^2F_1^d(x,Q^2)].
\end{equation}
The structure functions $F_{1,2}^d$ are related to the structure functions
$W_{1,2}$ (introduced in the formula (7)) by the following way:
$W_1=2F_1^d, \ W_2=4(\tau /y)F_2^d.$
The deuteron spin--independent structure function
$F_2^d=F_2^p(1+F_2^n/F_2^p)/2$ is calculated using
parameterizations for the proton structure functions $F_2^p$
\cite{ALLM97} and the ratio $F_2^n/F_2^p$ \cite{NMC}. The deuteron
spin--dependent structure function $b_2$ has also been extracted
from the experiment using the Callan--Gross relation
\begin{equation}\label{59}
b_2=2x\frac{1+R}{1+\gamma ^2}b_1.
\end{equation}

According to the preliminary results of the HERMES experiment the tensor
asymmetry can be parametrized as \cite{C1}
\begin{equation}\label{60}
A_{zz}=-1.56\cdot 10^{-2}(1-1.74x-1.45\sqrt{x}).
\end{equation}

\begin{figure}[ht]

\parbox{1.0\textwidth}{
\parbox{0.5\textwidth}{
\includegraphics[width=0.4\textwidth,height=0.42\textwidth]{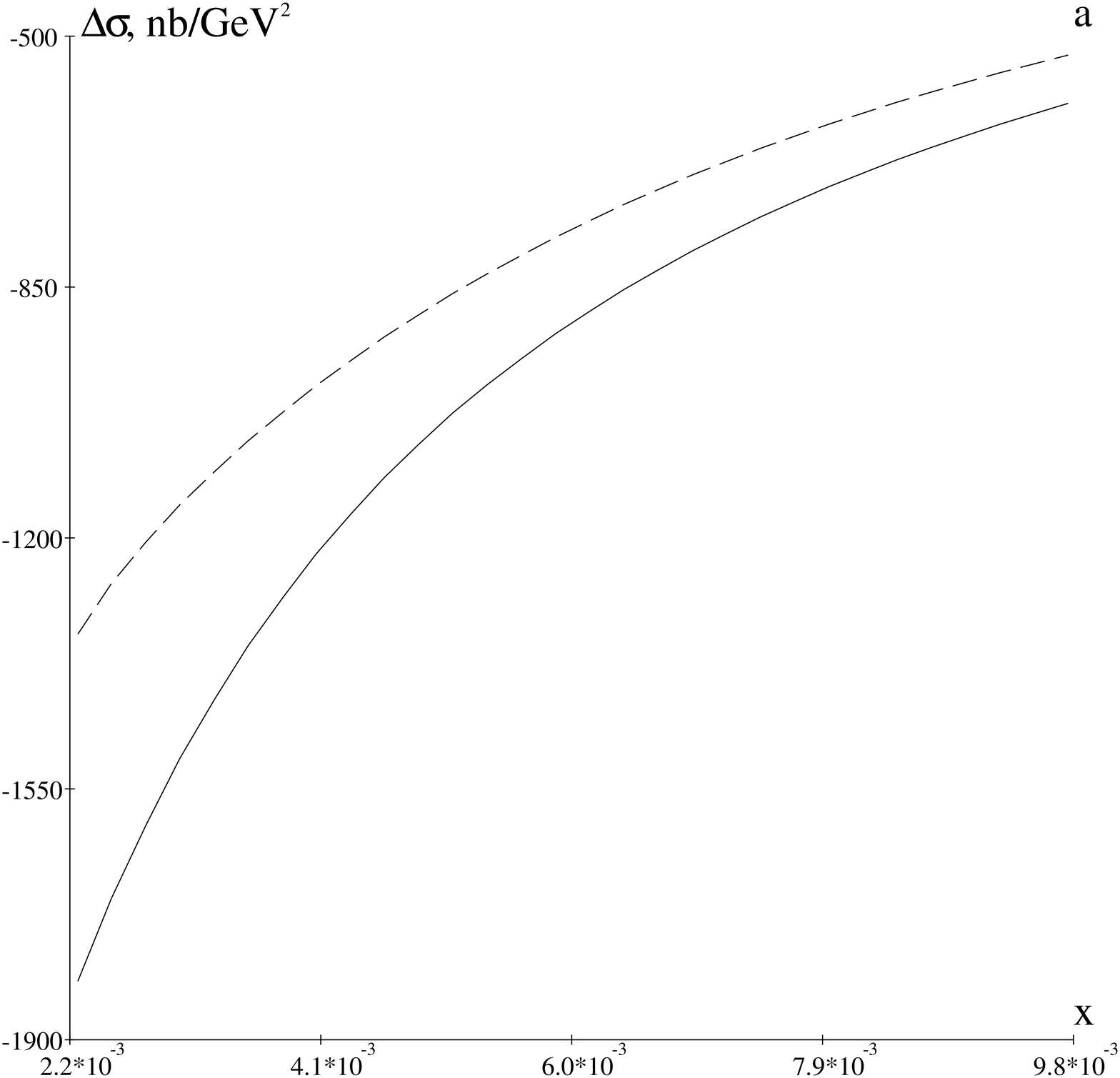}
\includegraphics[width=0.4\textwidth,height=0.42\textwidth]{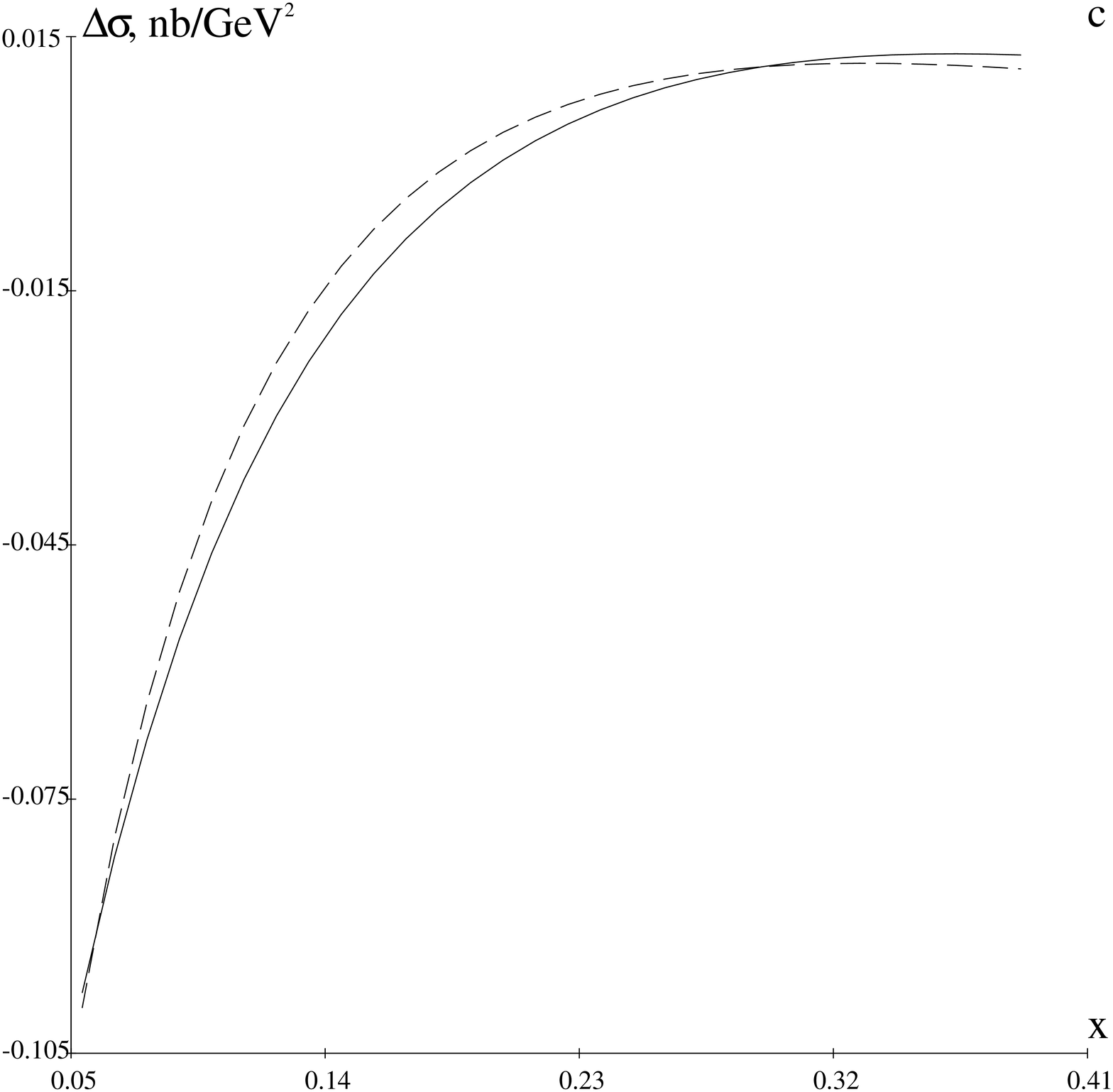}
}
\parbox{0.5\textwidth}{
\includegraphics[width=0.4\textwidth,height=0.42\textwidth]{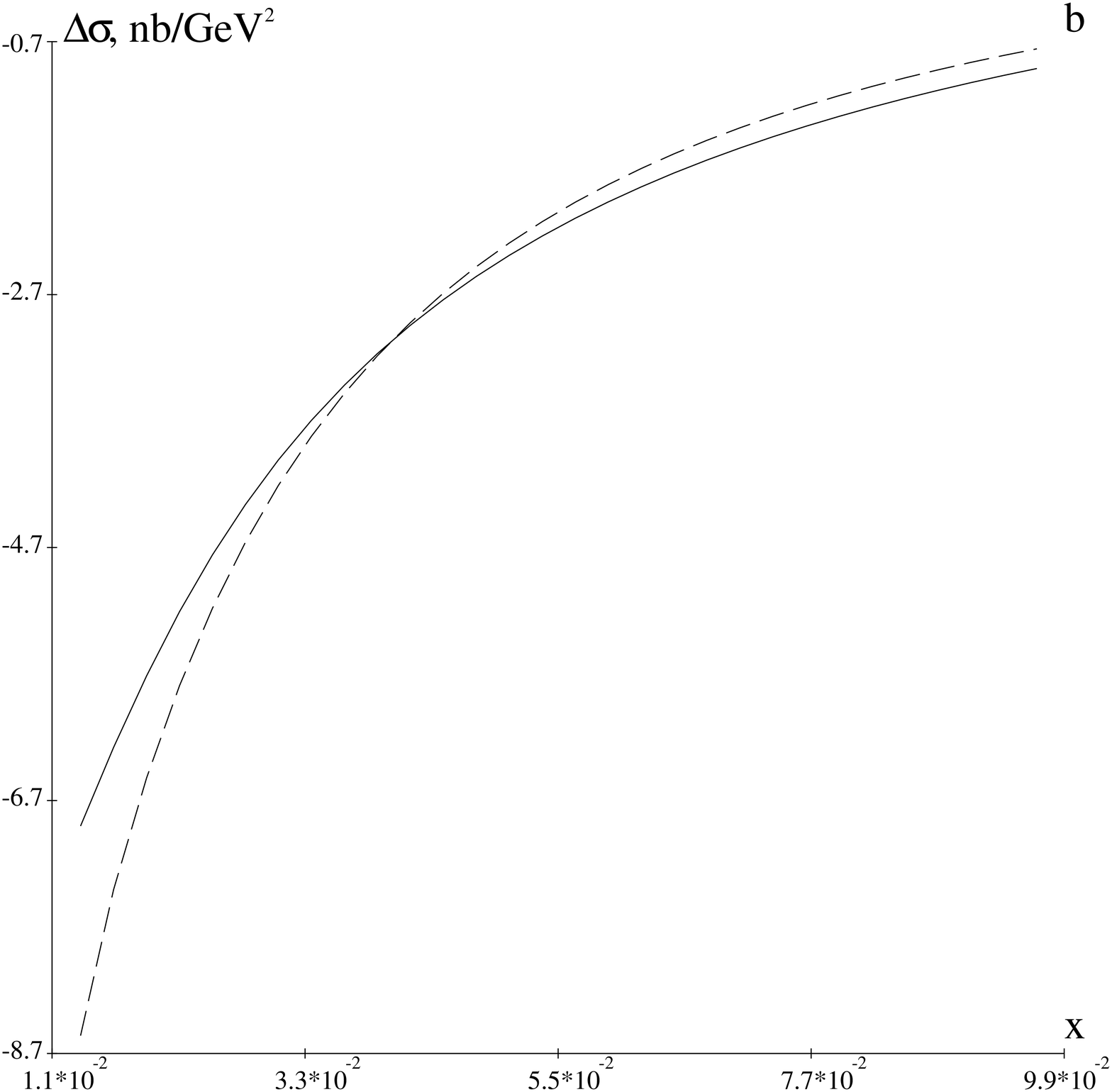}
\includegraphics[width=0.4\textwidth,height=0.42\textwidth]{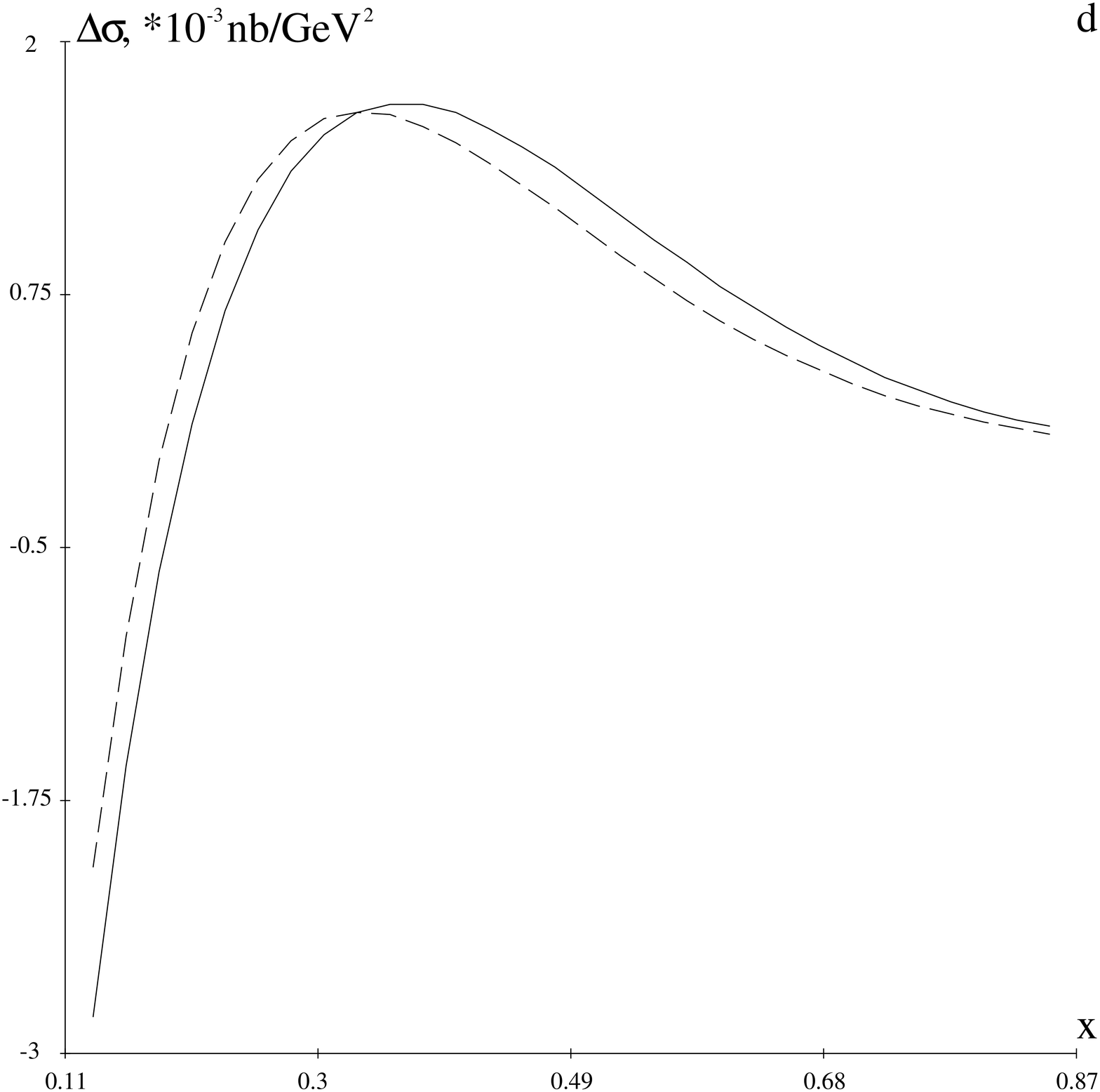}}}

\caption{The spin--dependent part of the cross section calculated
for the kinematical conditions of the HERMES experiment \cite{C}.
The solid line is the Born approximation, and the dotted line
corresponds to the inclusion of the radiative corrections. The
$Q^2$ values are the following: $a-0.1 GeV^2, b-1 GeV^2, c-4
GeV^2, d-10 GeV^2.$}
\end{figure}

The influence of the radiative correction on the spin--dependent
part of the Born cross--section  is shown in Fig.2 as a function
of the variable $x$ for various $Q^2$ values. The inclusion of the
radiative correction shifts the zero value of $b_1$ and $b_2$ to
the smaller x-value region (see Fig. 2 c and d). In the range of
low x ($x \sim 10^{-3}-10^{-2}$) the value of radiative correction
changes from $10\%$ to $30\%$ as compared with the Born
contribution. It is the region that  of the utmost importance for
$b_1$--measurements. According to the theoretical predictions
\cite{N,E,B} the structure function $b_1$ increases very rapidly
in this region and this fact was confirmed in the HERMES
experiment \cite{C}.

From our estimation we conclude that the radiative corrections to
process (1) are not small, especially for low $x$ region, and they
have to be taken into account in the data analysis.

\vspace{2cm}

{\bf Acknowledgments}

We wish to thank N. P. Merenkov for useful discussions and comments. We warmly
acknowledge M. Contalbrigo for useful discussions on the HERA experimental
conditions, as well as for sending us the preliminary results on the $A_{zz}$
parameterization.

\vspace{2cm}

\section{Appendix A}
\hspace{0.7cm}
\setcounter{equation}{0}
\def\theequation{A.\arabic{equation}}

In this Appendix we present the formulae for the coefficients $A^{mn}_i, \ $
$B^{mn}_i, \ $ and $C^{mn}_{ji}, $ ($m,n=l,t, \ i=1-4, \ j=0,1,2 $) that
determining the cross--section  of the hard--photon emission process (see
formula (50)).

The coefficients determining the contribution proportional to the $B_1$
structure function are:
$$A_1^{ll}=-\frac{n_1}{\tau }[(\bar r-\Delta_1)^2-2a(b+\Delta_1)], \
B_1^{ll}=\frac{n_1}{\tau }\Bigl\{[(2a-b)r+1+\Delta_2 )]^2-ar(2+3ar)\Bigr\}, \ $$
$$C_{01}^{ll}=-\frac{VN}{\tau }\Bigl\{(\bar r-\Delta_1)^2+
a[3a(1+r^2)-2(b+\Delta_1)]\Bigr\}, \  C_{11}^{ll}=-6N(c+2a), \
C_{21}^{ll}=-6N\frac{\tau }{V}, \ $$
\begin{equation}\label{A1}
A_{1}^{lt}=\frac{2n_1}{Md}Q_B^2(2b+\Delta_1)(\bar r-\Delta_1), \
B_{1}^{lt}=\frac{2n_1}{Md}Q_B^2(\Delta_2 -2br)[(a-b)r+1+\Delta_2 ], \
\end{equation}
$$C_{01}^{lt}=-4n_2Q_B^2\Bigl [a(1+r^2)(y+2a)-2b\bar r-
\Delta_1(c+2a-2b)\Bigr ], \
C_{11}^{lt}=-4n_2\Bigl [(y+4a)(c+2a)-2a(\bar r+2b)\Bigr ], \  $$
$$C_{21}^{lt}=-8n_2\frac{\tau }{V}(y+2a), \
A_{1}^{tt}=-\frac{n_1}{b}\frac{Q_B^2}{V}[b^2+(b+\Delta_1)^2], \
B_{1}^{tt}=\frac{n_1}{b}\frac{Q_B^2}{V}(2b^2r^2-2br\Delta_2 +\Delta_2^2), \  $$
$$C_{01}^{tt}=-\frac{NQ_B^2}{2b}\Bigl [(1+r^2)(y^2+4a-2ab)+
(2b+\Delta_1)^2+\Delta_1^2\Bigr ], \
C_{11}^{tt}=\frac{2N}{b}\Bigl [b(1+y+2a-r)+(1+a)\Delta_1\Bigr ], \  $$
$$C_{21}^{tt}=-\frac{N}{d^2}\Bigl [y^2+2a(2-b)\Bigr ]. \  $$
The coefficients determining the contribution proportional to the $B_2$
structure function are:
$$A_2^{ll}=\frac{n_3}{\tau }\Bigl [b(1+r^2)+(1-r+ry)\Delta_1\Bigr ]
\Bigl [(\bar r-\Delta_1)^2-2a(b+\Delta_1)\Bigr ], \ $$
$$B_2^{ll}=-\frac{n_3}{\tau }\Bigl [b(1+r^2)-\Delta_2 (\bar r-2a)\Bigr ]
\Bigl\{[(2a-b)r+1+\Delta_2 ]^2-ar(2+3ar)\Bigr\}, \ $$
$$C_{02}^{ll}=-\frac{NV}{c}\Bigl\{(7-3y)c^2+3a(5-y+r)c+3a^2(3+r^2)
-ar[5+3(a+b)^2]\Bigr\}, \ $$
$$C_{12}^{ll}=-3\tau \frac{N}{c}\Bigl [4(a+c)-cy\Bigr ], \
C_{22}^{ll}=-6N\frac{\tau ^2}{cV}, \ $$
$$A_2^{lt}=-2\frac{n_3Q_B^2}{Md}(2b+\Delta_1)(\bar r-\Delta_1)
\Bigl [b(1+r^2)+(1-r+ry)\Delta_1\Bigr ], \ $$
\begin{equation}\label{A2}
B_2^{lt}=-2\frac{n_3Q_B^2}{Md}(\Delta_2 -2br) [1+\Delta_2 +(a-b)r]
\Bigl [b(1+r^2)+(a+b-r)\Delta_2\Bigr ], \
\end{equation}
$$C_{02}^{lt}=\frac{NV^2}{Mdc}\Bigl\{2ac\Bigl [y\bar r+
(3b+a)(1+r)-y-8a\Bigr ]-c^2\Bigl [2a+(2-y)(y+4a)\Bigr ]+
2a\Bigl [2a(b-a+r)+(y+2a)(r- $$
$$-a(1+r^2)+r(a+b)^2)\Bigr ]\Bigr\}, \
C_{12}^{lt}=-4n_2\frac{\tau }{c}\Bigl [2a(1-3r+2\bar r+4c)+
cy(2+b-a)\Bigr ], \ C_{22}^{lt}=-8n_2\frac{\tau ^2}{cV}(y+2a), \ $$
$$A_2^{tt}=\frac{xy}{b}n_3Z_1[2b^2+\Delta_1(2b+\Delta_1)], \
B_2^{tt}=-\frac{xy}{b}n_3Z_2[2b^2r^2+\Delta_2 (\Delta_2 -2br)], \  $$
$$C_{02}^{tt}=\frac{VN}{2bc}\Bigl\{-2c^2\Bigl [a+(1+a)(2-y)\Bigr ]+
c\Bigl [(3-2y+a^2+b^2)(\bar r-2a)+4(ab+b-a^2)+4r(a-b^2)\Bigr ]- $$
$$-2a\Bigl [(r-a)^2+b^2\Bigr ]+(1+2a-2b+a^2+b^2)\Bigl [
r-a(1+r^2)+(a+b)^2r\Bigr ]\Bigr\}, \ $$
$$C_{12}^{tt}=-\frac{VN}{2cd^2}\Bigl\{c\Bigl [1+7a(1+a)-b(1+b)+
(a+b)(a^2+b^2)\Bigr ]+
4a\Bigl [(a-b)(1-r)+a^2+b^2-r\Bigr ]\Bigr\}, \ $$
$$C_{22}^{tt}=-\frac{N\tau }{cd^2}\Bigl [y^2+2a(2-b)\Bigr ]. $$
The coefficients determining the contribution proportional to the $B_3$
structure function are:
$$A_{3}^{ll}=n_3\frac{c}{\tau }\Bigl\{(a+\bar r)\Bigl [2Z_1+
r\Delta_1(2a+r-\Delta_1)\Bigr ]-\Delta_1
\Bigl [r^2(r-\Delta_1)+2(b+\Delta_1)+r(a+b)(a+r-\Delta_1)\Bigr ]\Bigr\}, \ $$
$$B_{3}^{ll}=-n_3\frac{c}{\tau }\Bigl\{2Z_2\Bigl [1+(2a-b)r+
\Delta_2 \Bigr ]+3a\Delta_2 \Bigl [(b-a)r-1-\Delta_2 \Bigr ]\Bigr\}, \ $$
\begin{equation}\label{A3}
C_{03}^{ll}=\frac{VN}{2}\Bigl [c(6a-16+9y)+6a(y-3-r)\Bigr ], \
C_{13}^{ll}=-3\tau N(2-y), \ C_{23}^{ll}=0,
\end{equation}
$$A_{3}^{lt}=-n_3\frac{cQ_B^2}{Md}\Bigl\{2Z_1(3b-a-r)+\Delta_1
\Bigl [4r(1+b^2+3ab)-2a(1+r^2)+c(ar-3+5br)\Bigr ]\Bigr\}, \ $$
$$B_{3}^{lt}=n_3\frac{cQ_B^2}{Md}\Bigl\{2Z_2\Bigl [(3b-a)r-1\Bigr ]
+\Delta_2 \Bigl [(1+ar)(a-6b)-(a+3b)(r^2+\Delta_2 )+r(b^2-1)+b+ $$
$$+\Delta_2 (3r-2b)\Bigr ]\Bigr\}, \
C_{03}^{lt}=n_2V\Bigl\{4a\Bigl [2b\bar r-y^2+4(b^2-a)\Bigr ]-
c\Bigl [3y(2-y)+8a(1+a+2b)\Bigr ]\Bigr\}, \ $$
$$C_{13}^{lt}=-2N\frac{M}{d}(2-y)(y+2a), \  C_{23}^{lt}=0,  $$
$$A_{3}^{tt}=n_3xy\frac{c}{b}\Bigl\{(\Delta_1-2b)[b(1+r^2)+(1-r+ry)
\Delta_1]+b\Delta_1[1+r(b-a+\Delta_1)]\Bigr\}, $$
$$B_{3}^{tt}=n_3xy\frac{c}{b}\Bigl\{(2br-\Delta_2)[b(1+r^2)+(1-r-y)
\Delta_2]+b\Delta_2[r(b-a+r)-\Delta_2)]\Bigr\}, $$
$$C_{03}^{tt}=\frac{VN}{2b}\Bigl\{3b-a-(a^2+b^2)(2+a+b)+r[y^2+2y(2b-a)+
2a(3-a)]+c[y(1+y+3a)-4(1+a)-2ab]\Bigr\}, $$
$$C_{13}^{tt}=-\frac{VN}{2d^2}(2-y)[y^2+2a(2-b)], \  C_{23}^{tt}=0.  $$
The coefficients determining the contribution proportional to the $B_4$
structure function are:
$$A_{4}^{ll}=n_3\frac{c^2}{\tau }\Bigl\{
(b-a)(1+r^2)+\Delta_1\Bigl [1+r(2a-b)\Bigr ]\Bigr\}, \
B_{4}^{ll}=n_3\frac{c^2}{\tau }\Bigl [
(a-b)(1+r^2)+\Delta_2 (a+\bar r)\Bigr ], \  $$
$$C_{04}^{ll}=-\frac{cNV}{2}\Bigl [1+3(b-a)\Bigr ], \
C_{14}^{ll}=C_{24}^{ll}=0, \ $$
$$A_{4}^{lt}=n_3\frac{c^2Q_B^2}{Md}\Bigl\{
2b(1+r^2)+\Delta_1\Bigl [1-r(3b-a)\Bigr ]\Bigr\}, \
B_{4}^{lt}=n_3\frac{c^2Q_B^2}{Md}\Bigl [
-2b(1+r^2)+\Delta_2 (\bar r-2b)\Bigr ], \  $$
\begin{equation}\label{A4}
C_{04}^{lt}=-c\frac{NV^2}{2Md}\Bigl [1+4ab-(a-b)^2\Bigr ], \
C_{14}^{lt}=C_{24}^{lt}=0, \
\end{equation}
$$A_{4}^{tt}=-xyrc^2\Delta_1n_3, \
B_{4}^{tt}=-xyc^2\Delta_2 n_3, \
C_{04}^{tt}=-\frac{cNV}{2}(y+2a), \
C_{14}^{tt}=C_{24}^{tt}=0, \ $$
here we use the following notation
$$c=z+xyr, \ \bar r=a-b+r, \ n_1=\frac{N}{2}\frac{1+r^2}{1-r}VQ_B^2, \
n_2=\frac{NV}{2Md}, \ n_3=\frac{N}{2c}\frac{V^2}{1-r}, \ $$
$$d^2=bQ_B^2, \ \Delta_1=(1-xy)r-a-b-z, \ \Delta_2=(1-y+xy)r+z-1, \
N=\frac{4\tau }{Vc^2}, $$
$$Z_1=b(1+r^2)+\Delta_1(1-r+yr), \ Z_2=b(1+r^2)+\Delta_2 (1-y-r). $$

\section{Appendix B}
\hspace{0.7cm}
\setcounter{equation}{0}
\def\theequation{B.\arabic{equation}}

In this Appendix we consider the part of the integral $I$ which is caused by
the $R_{ln}$ and $R_{tn}$ components of the deuteron quadrupole--polarization
tensor (these components do not contribute to the differential cross section
treated in the Born approximation). Let us define the integral which is caused
by the $R_{ln}$ component
\begin{equation}\label{B1}
I_{ln}=\int \frac{d^3{\vec k}}{2\pi\omega}\Sigma_{ln}(z,r,\varphi )R_{ln}, \
\end{equation}
with
$$\Sigma_{ln}(z,r,\varphi )=\frac{\alpha^2(q^2)}{Q_B^4}\frac{2VN}{Mr^2}n\cdot q
\Bigl (\frac{P_{1ln}}{\chi_1}-\frac{P_{2ln}}{\chi_2}+U_{0ln}+
U_{1ln}\chi_1\Bigr ), $$
$$P_{1ln}=\frac{V}{1-r}\Bigl\{cgxy(1+r^2)B_1+g\Bigl [c(1-r(1-y))+a(1+r^2)
-4fr\Bigr ]B_2+\Bigl [2a-fr+ $$
\begin{equation}\label{B2}
+\frac{1}{2}(3(1-r+yr)+2ar)\Bigr ]B_3+\frac{1}{2}c[1+(a-b)r]B_4\Bigr\},
\end{equation}
$$P_{2ln}=-\frac{V}{1-r}\Bigl\{-xy(1+r^2)(c+2ar)B_1+\frac{2}{c}
[-ar(a(1+r^2)-4fr)+\frac{c^2}{2}(1-y-r)+ $$
$$+\frac{c}{2}(4fr-a(1+3r^2+
2yr-2r))]B_2+[r(f-2ar)-\frac{c}{2}(2a+3(r+y-1))]B_3
-\frac{c}{2}(a-b+r)B_4\Bigr\}, $$
$$U_{0ln}=2g(cB_1+\tau B_2)+2\tau (2-y)(B_2+B_3), \
U_{1ln}=\frac{4\tau }{V}(B_1+\frac{\tau }{c}B_2), $$
and $c=z+xyr, \ f=1+(1-y)^2, \ g=1+2a/c, \ n\cdot q=S_{\mu}^{(n)}q_{\mu}.$

The second integral, caused by the $R_{tn}$ component, is defined as
\begin{equation}\label{B3}
I_{tn}=\int \frac{d^3{\vec k}}{2\pi\omega}\Sigma_{tn}(z,r,\varphi )R_{tn}, \
\end{equation}
where the integrand is
$$\Sigma_{tn}(z,r,\varphi )=\frac{\alpha^2(q^2)}{Q_B^4}\frac{2VN}{dr^2(r-1)}n\cdot q
\Bigl (\frac{P_{1tn}}{\chi_1}-\frac{P_{2tn}}{\chi_2}+U_{0tn}+
U_{1tn}\chi_1\Bigr ), $$
$$P_{1tn}=Q_B^2\Bigl\{\bar fxy(1+r^2)B_1-\frac{\bar f}{c}\Bigl [a(1+r^2)
-4r(f+4y)\Bigr ]B_2+ $$
$$+\bar f[1+r(y-1)](B_2+B_3)+brc(B_3+B_4)+2br(1-y)B_3\Bigr\}, $$
$$P_{2tn}=-Q_B^2\Bigl\{xy\bar g(1+r^2)B_1+\frac{\bar g}{c}
[a(1+r^2)-4fr]B_2+ $$
\begin{equation}\label{B4}
+\bar g(r-1+y)(B_2+B_3)-bc(B_3+B_4)-2br(1-y)B_3\Bigr\},
\end{equation}
$$U_{0tn}=(r-1)\Bigl [-2xy\bar f(B_1+\frac{\tau }{c}B_2)+
(2-y)(2a+y)(B_2+B_3)\Bigr ], \ $$
$$U_{1tn}=\frac{1}{V}\Bigl\{(r-1)(2a+y)G_1-\bar fG_2-\frac{y-2}{Mxy}
[M(2a+y)-2\tau d](B_2+B_3)-\frac{d}{Mxy}[c-2a(r-2)]G_2\Bigr\}, $$
$$G_1=3B_1+\frac{2\tau }{c}B_2+\frac{c}{2xyr}(B_2+2B_3+B_4), \
G_2=-B_1-\frac{c}{2xyr}(B_2+2B_3+B_4), $$
and $d^2=bQ_B^2, \ \bar f=b-a-z+r(1-xy), \ \bar g=z-1+r(a-b+xy).$

As before, we calculate the above integrals in c.m.s. of the hard photon and undetected
hadron system: ${\vec k_1}-{\vec k_2}+{\vec p}=0.$ The electron momenta
${\vec k_1}$ and ${\vec k_2}$ define the $xz$ plane, $z$ axis is directed
along the deuteron momentum ${\vec p}.$ Then the hard--photon momentum
${\vec k}$ is determined by the azimuthal ($\varphi $) and polar ($\theta $)
angles, and the phase space of the hard photon can be written as
\begin{equation}\label{B5}
\frac{d^3{\vec k}}{2\pi\omega}=\frac{Q_B^2}{2\sqrt{y^2+4a}}
\frac{d\varphi }{2\pi}dzdr,
\end{equation}
where $\omega $ is the hard--photon energy.

The quantity $n\cdot q$ can be written in this coordinate system as $n\cdot q
=\bar n sin\varphi ,$where $\bar n$ is a factor independent on $\varphi .$
Then the integration over the $\varphi $ variable in region $(0,2\pi)$ leads
to the following result: $I_{ln}=I_{tn}=0.$ So, the $R_{ln}$ and $R_{tn}$
components of the deuteron quadrupole--polarization tensor do not contribute
to the differential cross section of deep--inelastic scattering of unpolarized electron beam by the
tensor polarized target. This result is due to the fact that only the
scattered-electron variables are measured (it corresponds to the HERA
experimental conditions, for example).

If the hard--photon is detected then the $I_{ln}$ and $I_{tn}$ survive and
the expressions for $\Sigma_{ln}$ and $\Sigma_{tn}$ have to be taken into
account

\section{Appendix C}
\hspace{0.7cm}
\setcounter{equation}{0}
\def\theequation{C.\arabic{equation}}

In this Appendix we give some formulae describing the polarization state
of the deuteron target for different cases. For the case of arbitrary
polarization of the target it is described by the general spin--density
matrix (in general case it is defined by 8 parameters) which in the
coordinate representation has the form
\begin{equation}\label{C1}
\rho_{\mu\nu}=-\frac{1}{3}\bigl(g_{\mu\nu}-\frac{p_{\mu}p_{\nu}}{M^2}\bigr)
-\frac{i}{2M}\varepsilon_{\mu\nu\lambda\rho}s_{\lambda}p_{\rho}+ Q_{\mu\nu},
\ \ Q_{\mu\nu}=Q_{\nu\mu}, \ \ Q_{\mu\mu}=0\ , \ \ p_{\mu}Q_{\mu\nu}=0\ ,
\end{equation}
where $p_{\mu }$ is the deuteron 4-momentum, $s_{\mu}$ and $Q_{\mu\nu}$ are
the deuteron polarization 4-vector and the deuteron quadrupole--polarization
tensor.

In the deuteron rest frame the above formula is written as
\begin{equation}\label{C2}
\rho_{ij}=\frac{1}{3}\delta_{ij}+\frac{i}{2}\varepsilon
_{ijk}s_k+Q_{ij}, \ ij=x,y,z.
\end{equation}
This spin--density matrix can be written in the helicity representation
using the following relation
\begin{equation}\label{C3}
\rho_{\lambda\lambda'}=\rho_{ij}e_i^{(\lambda )*}e_j^{(\lambda')}, \
\lambda ,\lambda'=+,-,0,
\end{equation}
where $e_i^{(\lambda )}$ are the deuteron spin functions which have the
deuteron spin projection $\lambda $ on to the quantization axis (z axis).
They are
\begin{equation}\label{C4}
e^{(\pm )}=\mp \frac{1}{\sqrt{2}}(1,\pm i,0), \
e^{(0)}=(0,0,1).
\end{equation}
The elements of the spin--density matrix in the helicity representation
are related to the ones in the coordinate representation by such a way
\begin{equation}\label{C5}
\rho _{\pm\pm}=\frac{1}{3}\mp \frac{1}{2}s_z-\frac{1}{2}Q_{zz}, \
\rho_{00}=\frac{1}{3}+Q_{zz}, \
\rho_{+-}=-\frac{1}{2}(Q_{xx}-Q_{yy})+iQ_{xy}, \
\end{equation}
$$\rho_{+0}=-\frac{1}{2\sqrt{2}}(s_x-is_y)-
\frac{1}{\sqrt{2}}(Q_{xz}-iQ_{yz}),
\rho_{-0}=-\frac{1}{2\sqrt{2}}(s_x+is_y)+
\frac{1}{\sqrt{2}}(Q_{xz}+iQ_{yz}), \ \rho_{\lambda\lambda'}=
(\rho_{\lambda'\lambda})^* .  \  $$
To obtain these relations we use $Q_{xx}+Q_{yy}+Q_{zz}=0.$

The polarized deuteron target which is described by the population numbers
$n_+, \ $ $n_- $ and $n_0 $ is often used in the spin experiments. Here
$n_+, \ $ $n_- $ and $n_0 $ are the fractions of the atoms with
the nuclear spin projection on to the quantization axis
$m=+1, \ $ $m=-1$ and $m=0,$
respectively. If the spin--density matrix is normalizd to 1, i.e.
$Sp\rho =1$, then we have $n_++n_-+n_0=1.$ Thus, the polarization state of
the deuteron target is defined in this case by two parameters: the so--called
V (vector) and T (tensor) polarizations
\begin{equation}\label{C6}
V=n_+-n_-, \ T=1-3n_0.
\end{equation}
Using the definitions for the quantities $n_{\pm ,0}$
\begin{equation}\label{C7}
n_{\pm }=\rho_{ij}e_i^{(\pm )*}e_j^{(\pm )}, \
n_0=\rho_{ij}e_i^{(0)*}e_j^{(0)},
\end{equation}
we have the following relation between $V$ and $T$ parameters and parameters
of the spin--density matrix in the coordinate representation (in the case
when the quantization axis is directed along the z axis)
\begin{equation}\label{C8}
n_0=\frac{1}{3}+Q_{zz}, \ n_{\pm }=\frac{1}{3}\mp \frac{1}{2}s_z-
\frac{1}{2}Q_{zz},
\end{equation}
or
\begin{equation}\label{C9}
T=-3Q_{zz}, \ V=-s_z.
\end{equation}

\end{document}